\documentclass[prd,a4paper]{revtex4}

\usepackage{graphicx}
\usepackage{psfrag}
\usepackage{inputenc}
\usepackage{amsmath}
\usepackage{amssymb}
\usepackage{subfigure}

\inputencoding{latin9}

\newcommand{\nc}{\newcommand}

\def\lsim{\; \raise0.3ex\hbox{$<$\kern-0.75em
      \raise-1.1ex\hbox{$\sim$}}\; }
\def\gsim{\; \raise0.3ex\hbox{$>$\kern-0.75em
      \raise-1.1\textmd{}ex\hbox{$\sim$}}\; }

\nc{\be}[1]{\begin{equation}\mbox{$\label{#1}$}}
\nc{\bea}[1]{\begin{eqnarray} \mbox{$\label{#1}$}}
\nc{\Section}[2]{\section{#2}\label{#1}}
\nc{\Bibitem}[1]{\bibitem{#1}}
\nc{\Label}[1]{\label{#1}}
\nc{\ie}{{\em i.e. }}
\nc{\eg}{{\em e.g. }}
\nc{\eea}{\end{eqnarray}}
\nc{\ee}{\end{equation}}
\nc{\w}{\omega}
\begin{document}

\title{Non-Gaussianity in three fluid curvaton model}

\author{T. Multam\"aki}
\thanks{tuomul@utu.fi}
\affiliation{Department of Physics,
University of Turku, FIN-20014 Turku, FINLAND}
\author{J. Sainio}
\thanks{jtksai@utu.fi}
\affiliation{Department of Physics,
University of Turku, FIN-20014 Turku, FINLAND}
\author{I. Vilja}
\thanks{vilja@utu.fi}
\affiliation{Department of Physics,
University of Turku, FIN-20014 Turku, FINLAND}
\date{\today}

\begin{abstract}
The generation of non-gaussianity is studied in a three fluid curvaton model.
By utilizing second order perturbation theory we derive general formulae for the
large scale temperature fluctuation and non-gaussianity parameter, $f_{NL}$,
that includes the possibility of a non-adiabatic final state.
In the adiabatic limit we recover previously known results.
The results are applied to a three fluid curvaton model where the curvaton decays into
radiation and matter. We find that the amount of non-gaussianity decreases as the
final state of the system becomes more adiabatic and that the
generated non-gaussianity in the scenario is small, $|f_{NL}| \sim \mathcal{O}(1)$.

\end{abstract}

\maketitle

\section{Introduction}

The Cosmic Microwave Background (CMB) has proven to be a feast of information for modern cosmology. 
Ever since its discovery by Penzias and Wilson in 1964 \cite{Penzias:1965wn}, there have been a number of different experiments starting from RELIKT-1 \cite{Strukov:1984ab} resulting in drastic improvements to the quality of CMB data. A major scientific breakthrough was reached in 1992 when the COBE-satellite was able to detect the precense of the anisotropy in the CMB \cite{Smoot:1992td,Bennett:1996ce,Gorski:1996cf}.

The most recent data gathered by WMAP \cite{Komatsu:2003fd,Spergel:2006hy} is consistent with the hyphothesis that these perturbations were generated in an era of cosmic inflation. This is generally achieved with a slowly rolling scalar field which leads to an exponential expansion of the universe and the observed anisotropy is generated by the fluctuations of this inflaton field. Such a minimal scenario leads to adiabatic and gaussian perturbations in accordance with the current data.

A well motivated alternative to the simplest inflationary scenario is the curvaton mechanism \cite{Enqvist:2001zp,Lyth:2001nq,Moroi:2001ct,Assadullahi:2007uw,Bartolo:2002vf,Moroi:2002rd,Fujii:2002yx,Lyth:2002my,Sloth:2002xn,Hebecker:2002xw,Hofmann:2002gy,Moroi:2002vx,Enqvist:2002rf,Postma:2002et,Feng:2002nb,Gordon:2002gv,Dimopoulos:2002hm,Liddle:2003zw,Dimopoulos:2003ii,Enqvist:2003mr,Lyth:2003ip,Dimopoulos:2003ss,Dimopoulos:2003az,Kasuya:2003va,Endo:2003fr,Hamaguchi:2003dc,Bartolo:2003jx,Giovannini:2003mb,McDonald:2003jk,Chun:2004gx} in which the perturbations are generated by a second scalar field dubbed the curvaton which stays subdominant during inflation and the actual expansion of space is still driven by the inflaton. This allows the inflation potential to have more natural properties compared to the single field scenario and can still lead to adiabatic perturbations. However, the extra degrees of freedom in the system now allow for the possibility that the final state is not necessarily purely adiabatic.
Instead, the generation of an observable amount of isocurvature perturbations is a possibility that
can distinguish the curvaton scenario from the simple single field inflationary model.

Another ingredient that can differentiate the curvaton scenario from the simple inflaton hyphothesis is the concept of non-gaussianity, which has become more relevant with the data gathered by WMAP and with the advent of the Planck satellite. The current limits from the WMAP 5-year data state that the local non-gaussianity parameter $f_{NL}$ is limited to the values $-9 < f_{NL}^{local} < 111$ \cite{Komatsu:2008hk}. The single field inflationary scenario produces very little of non-gaussianity, $f_{NL}\sim\mathcal{O}(1)$ whereas the curvaton scenario might lead to an observable non-gaussianity.

The presence of non-gaussianity is therefore of paramount importance since it can differentiate between different scenarios of the early universe. This paper is focused closely on the generation of non-gaussianity in a three-fluid model of curvaton decay where the curvaton decays into both radiation and matter. The concept of non-gaussianity in this scenario was first presented in \cite{Gupta:2003jc} through first order perturbation theory. The study of non-gaussianities is however essentially dependent on the second-order terms and therefore the use of first order theory is not always justified. Our calculations incorporate the second-order perturbations from the start.
Since the curvaton scenario might lead to adiabatic and isocurvature perturbations we have derived a generalized non-gaussianity parameter that includes the adiabatic state as a special case.

This paper is organized as follows. In section II we present the relevant quantities and equations of motion of the perturbation up to second order. In this section we also derive the generalized non-gaussianity parameter $f_{NL}$. In section III we present the equations of motion of the curvaton model and generalize the conserved quantity first presented in \cite{Gupta:2003jc}. In section IV we present numerical results 
for $f_{NL}$ in the three fluid curvaton model both for the case of constant and time-dependent interaction.
We end this article with discussion and conclusions in section V.

\section{Perturbations at first and second order}

The theory of second order cosmological perturbations has been studied rigorously in the recent years. We will follow closely the notation of \cite{Bartolo:2004if} and use a spatially flat Friedmann-Robertson-Walker -background. The metric tensor can be in this case expanded 
up to second order in the form \cite{Bartolo:2004if}
\be{metric}
\begin{aligned}
g_{\mu\nu}dx^{\mu}dx^{\nu} = & -(1+2\phi^{(1)}+\phi^{(2)})dt^2+a(t)(\hat{\omega}^{(1)}_{i}+\frac{1}{2}\hat{\omega}^{(2)}_{i})dtdx^{i}\\
& +a(t)^2\Big[(1-2\psi^{(1)}-\psi^{(2)})\delta_{ij}+(\chi^{(1)}_{ij}+\frac{1}{2}\chi^{(2)}_{ij}\Big]dx^idx^j,
\end{aligned}
\ee
where $a(t)$ is the scale factor and $\phi^{(r)}$, $\hat{\omega}^{(r)}_{i}$, $\psi^{(r)}$ and $\chi^{(r)}_{ij}$ are  perturbation functions
defined in \cite{Bartolo:2004if} at first ($r=1$) and second order ($r=2$). Written in this form, different gauges can be 
straightforwardly given in terms of the perturbation functions: for example the Poisson gauge is defined as $\omega^{(r)} = \chi^{(r)}_{ij} = \chi^{(r)}_{ij} = 0$ and the spatially flat one is $\psi^{(r)} = \chi^{(r)} = 0$ \cite{Bartolo:2004if}.

A useful set of equations can be derived from the continuity equations $T^{\mu\nu}_{i;\mu}=Q^{\nu}_{i}$, where $T^{\mu\nu}_{i}$ is the energy-momentum tensor, $Q_{i}$ describes the energy transfer between different fluids and $;$ denotes the covariant derivative. 
From the continuity equations it follows that the equations determining the background evolution of individual fluids are
\begin{equation} \label{eq:conti}
\dot{\rho}_{i} = -3H(1+\omega_{i})\rho_{i} + Q_{i},
\end{equation}
where $\w_{i}=P_{i}/\rho_{i}$ is the equation of state of th $i$th fluid and $\dot{} \equiv d/dt$ \ie derivative
with respect to physical time.

At first order one finds the evolution equations of the perturbed energy and pressure densities
(on large scales) \cite{Malik:2004tf}
\be{fluidpert-1}
\dot{\delta\rho}^{(1)}_{i}+3H(\delta\rho^{(1)}_{i}+\delta P^{(1)}_{i})-3(\rho_{i}+P_{i})\dot{\psi}^{(1)} = Q_{i}\phi^{(1)} + \delta Q^{(1)}_{i}.
\ee
and at second order
\be{fluidpert-2}
\begin{aligned}
&\dot{\delta\rho}^{(2)}_{i}+3H(\delta\rho^{(2)}_{i}+\delta P^{(2)}_{i})-3(\rho_{i}+P_{i})\dot{\psi}^{(2)} - 6\dot{\psi}^{(1)}[\delta\rho_{i}+\delta P_{i}+2(\rho_{i}+P_{i})\psi^{(1)}] \\
&= Q_{i}\phi^{(2)} + \delta Q^{(2)}_{i} - Q_{i}(\phi^{(1)})^{2}+2\phi^{(1)}\delta Q^{(1)}_{i}.
\end{aligned}
\ee

In addition to these, we also have the Einstein equations which can be used to give additional limits on the perturbation equations. For example in the Poisson gauge at first order $\psi^{(1)}=\phi^{(1)}$ and on large-scales $2\psi^{(1)}=-\delta\rho/\rho_{0}$, whereas in the spatially flat gauge $\psi^{(1)}=\chi^{(1)}=0$ and $2\phi^{(1)}=-\delta\rho/\rho_{0}$ \cite{Bartolo:2004if}.

At second order the equations get more complex and they are presented in detail in \cite{Bartolo:2004if}. The equations which we will need 
are the $0-0$ and $i-j$-components of the Einstein equations on large scales which read in the Poisson gauge in a matter dominated universe as
\begin{equation} \label{eq:E-so}
\begin{aligned}
\phi^{(2)} = &-\frac{1}{2}\frac{\delta\rho^{(2)}}{\rho_{0}}+4(\psi^{(1)})^{2}\\
\psi^{(2)} - \phi^{(2)} = & -4(\psi^{(1)})^{2} - \frac{10}{3}\nabla^{-2}(\partial^{i}\psi^{(1)}\partial_{i}\psi^{(1)})\\
& +10 \nabla^{-4}\Big(\partial^{i}\partial_{j}\Big(\partial_{i}\psi^{(1)}\partial^{j}\psi^{(1)}\Big)\Big).\\
\end{aligned}
\end{equation}
In the spatially flat gauge the $0-0$-component is
\begin{equation}
\phi^{(2)} = -\frac{1}{2}\frac{\delta\rho^{(2)}}{\rho_{0}}+4(\phi^{(1)})^{2}.\\
\end{equation}

\subsection{Curvature perturbations}

An elegant way to study the evolution of perturbations is to use gauge-invariant curvature perturbations, which relate to curvature perturbations on homogeneous-density surfaces. At first order they are defined for component $i$ as
\begin{equation}
\zeta^{(1)}_{i}=-\psi^{(1)}-\frac{\delta\rho^{(1)}_{i}}{\rho_{i}'},
\end{equation}
where $'\equiv d/d(\ln(a))$. At second order the corresponding quantity is defined as
\begin{equation} \label{eq:zeta-2}
\zeta^{(2)}_{i}=-\psi^{(2)}-\frac{\delta\rho^{(2)}_{i}}{\rho_{i}'}+2\frac{\delta\rho^{(1)}_{i}{}'}{\rho_{i}'}\frac{\delta\rho^{(1)}_{i}}{\rho_{i}'}+2\frac{\delta\rho^{(1)}_{i}}{\rho_{i}'}(\psi^{(1)}{}'+2\psi^{(1)})-\Big(\frac{\delta\rho^{(1)}_{i}}{\rho_{i}'}\Big)^{2}\Big(\frac{\rho_{i}''}{\rho_{i}'}-2\Big).
\end{equation}
Note that we are here neglecting gradient terms since we are only interested in the large scale behaviour of perturbations.

The equation of motion of the first order curvature perturbations can be derived from eq. (\ref{fluidpert-1}) and Einstein equations. The result is
\begin{equation} \label{eq:zeta-eq-1}
\begin{aligned}
\zeta^{(1)}_{i}{}' = & \frac{3\delta P_{int(i)}}{\rho_{i}'} - \frac{\delta Q_{int (i)}}{H\rho_{i}'} -\frac{H'}{H}\frac{Q_{i}}{\rho_{i}'}(\zeta - \zeta_{i}),\\
\end{aligned}
\end{equation}
where $\delta P_{int(i)} \equiv \delta P_{i} - p_{i}'\delta \rho_{i}/\rho_{i}'$ and $\delta Q_{int (i)} \equiv \delta Q_{i}-Q'_{i}\delta \rho_{i}/\rho'_{i}$.

At second order the corresponding equations read as \cite{Bartolo:2003bz}
\begin{equation} \label{eq:zeta-eq-2}
\begin{aligned}
\zeta^{(2)}_{i}{}' = & -\frac{1}{\rho_{i}'H}\Big[\Big(\delta Q^{(2)}_{i}-\frac{Q_{i}'}{\rho_{i}'}\delta\rho^{(2)}_{i}\Big)+Q_{i}\frac{\rho_{0}'}{2\rho_{0}}\Big(\frac{\delta\rho^{(2)}_{i}}{\rho_{i}'}-\frac{\delta\rho^{(2)}}{\rho_{0}'}\Big)\Big]-3\frac{Q_{i}}{\rho_{i}'H}\Big(\phi^{(1)}\Big)^2\\
& -2\frac{\delta Q^{(1)}_{i}\phi^{(1)}}{H\rho_{i}'}+2\Big[2-3(1+\omega_{i})\Big]\zeta^{(1)}_{i}\zeta^{(1)}_{i}{}'-2\Big[\Big(\frac{Q_{i}\phi^{(1)}}{H\rho_{i}'}+\frac{\delta Q^{(1)}_{i}}{H\rho_{i}'}\Big)\zeta^{(1)}_{i}\Big]' - \Big[\Big(\frac{Q_{i}'}{H\rho_{i}'}-\frac{1}{2}\frac{Q_{i}}{H\rho_{i}'}\frac{\rho_{0}'}{\rho_{0}}\Big)\Big(\zeta^{(1)}_{i}\Big)^{2}\Big]'.
\end{aligned}
\end{equation}

There are instances when the definition of different curvature perturbations might fail \eg when $\rho_{i}'=0$. Therefore in our numerical evaluations we have used the spatially flat gauge and evaluated the density perturbations of different components in this gauge at first and second-order. The corresponding equations of motion can be easily read from equations (\ref{fluidpert-1}) and (\ref{fluidpert-2}) by going to the spatially flat gauge $\psi^{(r)} = \chi^{(r)} = 0$.


\subsection{Non-gaussianity}

The generation of non-gaussianity in the two-fluid curvaton model at second order has been considered previously in 
in \cite{Bartolo:2004if,Sasaki:2006kq}. In the two-fluid model the final state is adiabatic whereas in the
three fluid model a significant isocruvature component is a possibility. Here we generalize the
the results of \cite{Bartolo:2004if} to include the possibility of a non-adiabatic final state.
We follow the standard notation presented in \cite{Mollerach:1997up} and use the Poisson gauge. Since we are interested only in the large scale non-gaussianities we can safely ignore some terms from the full expression \cite{Mollerach:1997up}, including the integrated Sachs-Wolfe effect. The temperature fluctuations can in this approximation be written  \cite{Bartolo:2004if} as
\begin{equation} \label{eq:SW}
\frac{\Delta T}{T} = \Big[\phi^{(1)}+\tau^{(1)}+\frac{1}{2}\Big(\phi^{(2)} + \tau^{(2)}\Big) - \frac{1}{2}\Big(\phi^{(1)}\Big)^2 + \phi^{(1)}\tau^{(1)}\Big]_{Em},
\end{equation}
where $\phi^{(r)}$ are the metric lapse functions, $\tau = \tau^{(1)} + \frac{1}{2}\tau^{(2)}$ is the intrinsic fractional temperature fluctutation $\tau=\Delta T/T|_{Em}$ and all the terms are evaluated at the time of emission.

We introduce new variables relating the final values of perturbations to their initial values:
\begin{equation}
\begin{aligned}
r_{1} & = \frac{\zeta^{(1)}_m|_m}{\zeta^{(1)}_{\sigma,\textrm{in}}}, \quad q_{1}=\frac{\zeta^{(1)}_{\gamma}|_m}{\zeta^{(1)}_{\sigma,\textrm{in}}}\\
r_{2} & =\frac{\zeta^{(2)}_m|_m}{(\zeta^{(1)}_{\sigma,\textrm{in}})^2}, \quad q_{2}=\frac{\zeta^{(2)}_{\gamma}|_m}{(\zeta^{(1)}_{\sigma,\textrm{in}})^2},\\
\end{aligned}
\end{equation}
where the different numerators are evaluated at the time of decoupling, when the universe is matter dominated and $\zeta^{(i)}\simeq\zeta^{(i)}_{m}$. The system is adiabatic if $q_{1}=r_{1}$ and $q_{2}=r_{2}$.

Since the universe is matter dominated during decoupling we can write $\zeta_{m}^{(2)}$ in the form
\begin{equation}
\zeta^{(2)}_{m}=-\psi^{(2)}+\frac{1}{3}\frac{\delta\rho^{(2)}}{\rho_{0}}+\frac{5}{9}\Big(\frac{\delta\rho^{(1)}}{\rho_0}\Big)^{2}.
\end{equation}
From the definitions of $r_1$ and $r_2$ it follows that
\begin{equation}
\zeta^{(2)}_{m} = r_2(\zeta^{(1)}_{\sigma,\textrm{in}})^2=r_2\Big(\frac{\zeta^{(1)}_{m}}{r_1}\Big)^2=\frac{25r_2}{9r_1^2}(\psi^{(1)})^2,
\end{equation}
where we have also used the equation $\phi^{(1)}=-3\zeta^{(1)}/5$ which is valid on large-scales in a matter dominated universe. By combining these equations with the Einstein equations (\ref{eq:E-so}), we can write $\phi^{(2)}$ in a matter dominated universe in the form
\begin{equation}
\phi^{(2)} = \Big[\frac{16}{3}-\frac{5}{3}\frac{r_2}{r_1^2}\Big](\psi^{(1)})^2+2\nabla^{-2}\Big(\partial^{i}\psi^{(1)}\partial_{i}\psi^{(1)}\Big) - 6\nabla^{-4}\Big(\partial^{i}\partial_{j}\Big(\partial_{i}\psi^{(1)}\partial^{j}\psi^{(1)}\Big)\Big),
\end{equation}
where the inverse Laplacians are to be understood as a formal expression.

The intrinsic fractional temperature fluctuations $\tau^{(r)}$ can be written at first order as
\begin{equation}
\tau^{(1)}=\frac{1}{4}\frac{\delta\rho^{(1)}_{\gamma}}{\rho_{\gamma}}\Big|_{Em}=-\frac{\delta\rho^{(1)}_{\gamma}}{\rho_{\gamma}'}\Big|_{Em}=\psi^{(1)}\Big|_{Em}+\zeta^{(1)}_{\gamma}\Big|_{Em}=\Big(1-\frac{5}{3}\frac{q_1}{r_1}\Big)\phi^{(1)}\Big|_{Em},
\end{equation}
where we have used the definitions of $\zeta^{(1)}_{\gamma}$, $\zeta^{(1)}_{m}$, $q_1$, $r_1$ and $\phi^{(1)}=-3\zeta^{(1)}/5$. At second order the corresponding variable is
\begin{equation}
\tau^{(2)}=\frac{1}{4}\frac{\delta\rho^{(2)}_{\gamma}}{\rho_{\gamma}}\Big|_{Em}-3\Big(\tau^{(1)}\Big)^2.
\end{equation}
From the definition of $\zeta^{(2)}_{\gamma}$ we find
\begin{equation}
\frac{1}{4}\frac{\delta\rho^{(2)}_{\gamma}}{\rho_{\gamma}}\Big|_{Em}=\psi^{(2)}+2\Big(\tau^{(1)}\Big)+4\tau^{(1)}\phi^{(1)}+\frac{25}{9}\frac{q_2}{r_1^2}\Big(\phi^{(1)}\Big)^2,
\end{equation}
where we have also used the definitions of $r_1$ and $q_2$.

Combining all of the above expressions for perturbations and substituting into the equation for the temperature fluctuations (\ref{eq:SW}) one finally has
\begin{equation} \label{eq:DT}
\begin{aligned}
\frac{\Delta T}{T} = \frac{6r_1-5q_1}{3r_1}\Big[&\phi^{(1)}+\Big[\frac{25(q_2-q_1^2)-60q_1r_1+96r_1^2-30r_2}{6r_1(6r_1-5q_1)}\Big]\Big(\phi^{(1)}\Big)^2\\
& + \frac{r_1}{6r_1-5q_1}\nabla^{-2}\Big(\partial^{i}\psi^{(1)}\partial_{i}\psi^{(1)}\Big) - \frac{3r_1}{6r_1-5q_1}\nabla^{-4}\Big(\partial^{i}\partial_{j}\Big(\partial_{i}\psi^{(1)}\partial^{j}\psi^{(1)}\Big)\Big) \Big],\\
\end{aligned}
\end{equation}
where we have ignored the momentum dependent terms because we are interested in the large scale non-gaussianity \cite{Bartolo:2004if}. To our knowledge this general formula has not been presented before. From eq. (\ref{eq:DT}) we can read that in the adiabatic limit, \ie $r_1=q_1$ and $r_2=q_2$, we recover at second order an extension of the first order Sachs-Wolfe effect $\Delta T/T=\phi^{(1)}/3$ given in \cite{Bartolo:2004if}:
\begin{equation}
\frac{\Delta T}{T}=\frac{1}{3}\Big[\phi^{(1)}+\frac{1}{2}\Big(\phi^{(2)}-\frac{5}{3}\big(\phi^{(1)}\big)^2\Big) + \nabla^{-2}\Big(\partial^{i}\psi^{(1)}\partial_{i}\psi^{(1)}\Big) - 3\nabla^{-4}\Big(\partial^{i}\partial_{j}\Big(\partial_{i}\psi^{(1)}\partial^{j}\psi^{(1)}\Big)\Big)\Big].
\end{equation}
In the opposite case of pure isocurvature perturbation, equation (\ref{eq:DT}) is not valid since we have assumed that $r_1\neq0$ and for isocurvature perturbations $0=\zeta^{(1)}\Big|_{Dec}\simeq\zeta^{(1)}_m\Big|_{Dec}=r_1\zeta^{(1)}_{\sigma,\textrm{in}}$.

We can now define the non-gaussianity parameter $f_{NL}$ in the general case. Following the notation of \cite{Komatsu:2002db} we write the temperature fluctuations in the form
\begin{equation}
\frac{\Delta T}{T}=g_T\Big[\phi^{(1)}+f_{NL}\big(\phi^{(1)}\big)^2\Big],
\end{equation}
where the factor $g_T$ depends on the state of the system \eg for a completely adiabatic one $g_{T}=1/3$ and in our calculations $g_T=(6r_1-5q_1)/(3r_1)$. We have ignored the gradient terms from our definition of non-gaussianity but when calculating the bispectrum they need to be included. Note that this definition leads to a sign difference when compared to the usual approach using the Bardeen potential \cite{Bartolo:2004if}.
Now from equation (\ref{eq:DT}) we can easily read the non-gaussianity parameter
\begin{equation} \label{eq:fNL}
f_{NL}=\frac{25(q_2-q_1^2)-60q_1r_1+96r_1^2-30r_2}{6r_1(6r_1-5q_1)},
\end{equation}
which is well defined since $r_1\neq0$ and $q_1\leq r_1$.

In order to compare with observations we need to calculate the bispectrum, which requires that we take into account other effects \eg the integrated Sachs-Wolfe effect and possible evolution after horizon crossing.
This is beyond the scope of this paper and hence left for further work. For details of this procedure we refer to
\cite{Komatsu:2001rj, Bartolo:2005kv} where the radiation transfer functions depend now on the adiabicity of the
system via $g_T$.

\section{The curvaton model}

\subsection{Evolution equations}

We denote the curvaton by subscript ($\sigma$), radiation by ($\gamma$) and matter by ($m$). Since the curvaton field is oscillating it can be safely estimated \cite{Ferrer:2004nv} to behave like non-relativistic matter, \ie $\omega_{\sigma}=0$. For radiation and matter we have $\omega_{\gamma}=1/3$ and $\omega_{m}=0$ and the interaction terms are \cite{Gupta:2003jc}
\begin{equation} \label{eq:interact}
\begin{aligned}
Q_{\sigma} & = -\Gamma_{\gamma}f_{\gamma}(N)\rho_{\sigma} - \Gamma_{m}f_{m}(N)\rho_{\sigma}\\
Q_{\gamma} & = \Gamma_{\gamma}f_{\gamma}(N)\rho_{\sigma}\\
Q_{m} & = \Gamma_{m}f_{m}(N)\rho_{\sigma},\\
\end{aligned}
\end{equation}
where $\Gamma_{i}$ denote the strength of the interaction and functions $f_{i}(N)$ allow for time dependent interactions
with $N\equiv \ln(a)$.

Background equations (\ref{eq:conti}) can be written in terms of fractional densities $\Omega_{i} \equiv \rho_{i}/\rho$ for which the equations of motion are \cite{Gupta:2003jc}:
\begin{equation} \label{eq:bgr}
\begin{aligned}
\Omega_{\sigma}' & =\Omega_{\sigma}\Omega_{\gamma}+\frac{Q_{\sigma}}{H\rho} ,\\
\Omega_{\gamma}' & =\Omega_{\gamma}(\Omega_{\gamma}-1)+\frac{Q_{\gamma}}{H\rho} ,\\
\Omega_{m}' & =\Omega_{m}\Omega_{\sigma}+\frac{Q_{m}}{H\rho} ,\\
\Big(\frac{1}{H}\Big)' & = \Big(1+\frac{1}{3}\Omega_{\gamma}\Big)\Big(\frac{1}{H}\Big).\\
\end{aligned}
\end{equation}
From the definition of $\Omega_{i}$ it can be easily seen that $\Omega_{\sigma}+\Omega_{\gamma}+\Omega_{m}=1$, which means that one equation of motion of $\Omega_{i}$ is redundant.

From eq. (\ref{fluidpert-1}) we can read the equations of motion for the first order density perturbations
\begin{equation} \label{eq:flat-rho1}
\begin{aligned}
\delta\rho^{(1)}_{\sigma}{}' & = -3\delta\rho^{(1)}_{\sigma} - \frac{Q_{\sigma}}{H}\frac{\delta\rho^{(1)}}{2\rho} + \frac{\delta Q^{(1)}_{\sigma}}{H},\\
\delta\rho^{(1)}_{\gamma}{}' & = -4\delta\rho^{(1)}_{\gamma} - \frac{Q_{\gamma}}{H}\frac{\delta\rho^{(1)}}{2\rho} + \frac{\delta Q^{(1)}_{\gamma}}{H},\\
\delta\rho^{(1)}_{m}{}' & = -3\delta\rho^{(1)}_{m} - \frac{Q_{m}}{H}\frac{\delta\rho^{(1)}}{2\rho} + \frac{\delta Q^{(1)}_{m}}{H}\\
\end{aligned}
\end{equation}
in the spatially flat gauge. At second order the corresponding equations in the flat gauge are
\begin{equation} \label{eq:flat-rho2}
\begin{aligned}
\delta\rho^{(2)}_{\sigma}{}' & = -3\delta\rho^{(2)}_{\sigma} - \frac{Q_{\sigma}}{2H\rho}\Bigg[\delta\rho^{(2)} -\frac{3}{2}\frac{\Big(\delta\rho^{(1)}\Big)^2}{\rho}\Bigg] + \frac{\delta Q^{(2)}_{\sigma}}{H} - \frac{\delta\rho^{(1)}\delta Q^{(1)}_{\sigma}}{H\rho},\\
\delta\rho^{(2)}_{\gamma}{}' & = -4\delta\rho^{(2)}_{\gamma} - \frac{Q_{\gamma}}{2H\rho}\Bigg[\delta\rho^{(2)} -\frac{3}{2}\frac{\Big(\delta\rho^{(1)}\Big)^2}{\rho}\Bigg] + \frac{\delta Q^{(2)}_{\gamma}}{H} - \frac{\delta\rho^{(1)}\delta Q^{(1)}_{\gamma}}{H\rho},\\
\delta\rho^{(2)}_{m}{}' & = -3\delta\rho^{(2)}_{m} - \frac{Q_{m}}{2H\rho}\Bigg[\delta\rho^{(2)} -\frac{3}{2}\frac{\Big(\delta\rho^{(1)}\Big)^2}{\rho}\Bigg] + \frac{\delta Q^{(2)}_{m}}{H} - \frac{\delta\rho^{(1)}\delta Q^{(1)}_{m}}{H\rho}.\\
\end{aligned}
\end{equation}

We also need the gauge invariant curvature perturbations $\zeta^{(i)}_{j}$ in order to calculate the non-gaussianity parameter $f_{NL}$. In the spatially flat gauge these are
\begin{equation} \label{eq:zetas-flat}
\begin{aligned}
\zeta^{(1)}_{i} =&-\frac{\delta\rho^{(1)}_i}{\rho_{i}'},\\
\zeta^{(2)}_{i} =&-\frac{\delta\rho^{(2)}_i}{\rho_{i}'}+\Big[2-3(1+\omega_{i})\Big]\Big(\zeta^{(1)}_{i}\Big)^2 - 2\Big[\frac{Q_{i}\phi^{(1)}}{H\rho_{i}'}+\frac{\delta Q^{(1)}_{i}}{H\rho_{i}'}\Big]\zeta^{(1)}_i\\
&-\Big[\frac{Q_i'}{\rho_i'H}-\frac{1}{2}\frac{Q_i \rho'}{\rho_i'H\rho}\Big]\Big(\zeta^{(1)}_{i}\Big)^2.
\end{aligned}
\end{equation}

The set of equations (\ref{eq:bgr}), (\ref{eq:flat-rho1}) and (\ref{eq:flat-rho2}) can now be evaluated numerically once the initial values have been set. We have chosen the system to be initially radiation dominated and non-adiabatic at first and second-order. The energy density of the curvaton field is during oscillations $\rho_{\sigma}=m^2\sigma^{2}$ where $\sigma$ is the amplitude of the field. A small perturbation of the field $\delta\sigma$ leads to \cite{Bartolo:2003jx}
\begin{equation}
\tilde{\rho}_{\sigma} = \rho_{\sigma} + \delta\rho^{(1)}_{\sigma} + \frac{1}{2}\delta\rho^{(2)}_{\sigma} + \mathcal{O}(\delta\rho^{(3)}_{\sigma})= m^2\Big(\sigma^{2}_0 + 2\sigma_0\delta\sigma + (\delta\sigma)^{2}\Big),
\end{equation}
which means that at second order
\begin{equation} \label{eq:curv-dens-2}
\frac{\delta\rho^{(2)}_{\sigma}}{\rho_{\sigma}}=\frac{1}{2}\Big(\frac{\delta\rho^{(1)}_{\sigma}}{\rho_{\sigma}}\Big)^2.
\end{equation}
This forces the second order perturbation of the curvaton to be 
\begin{equation} \label{eq:zeta-2-curv-in}
\zeta^{(2)}_{\sigma,\textrm{in}}=\Big[\frac{1}{2}-\frac{\Gamma_{\sigma}}{2H}\Omega_{\gamma}-\frac{2\Gamma_{\sigma}\Omega_{\gamma}}{3H+\Gamma_{\sigma}}+\frac{(\Gamma_{\sigma})^2}{2H(3H+\Gamma_{\sigma})}\Big]\Big(\zeta^{(1)}_{\sigma}\Big)^2\Big|_{\textrm{in}},
\end{equation}
which follows from eqs. (\ref{eq:curv-dens-2}) and (\ref{eq:zeta-2}) after short calculations. Now since the first order curvature perturbations are linear equations we can scale them and set $\zeta^{(1)}_{\sigma,\textrm{in}}=1$. By pluggins this into eq. (\ref{eq:zeta-2-curv-in}) and using the inequality $\Gamma_{\sigma}\ll H_0$, valid in our calculations, we can safely estimate $\zeta^{(2)}_{\sigma,\textrm{in}}\simeq 1/2$. The other perturbations are initally set to be zero \ie $\zeta^{(1)}_{\gamma,\textrm{in}}=\zeta^{(1)}_{m,\textrm{in}}=\zeta^{(2)}_{\gamma,\textrm{in}}=\zeta^{(2)}_{m,\textrm{in}}=0$ since we have assumed that only the curvaton field has initial perturbations.
Based on these initial values and equations (\ref{eq:bgr}), (\ref{eq:flat-rho1}) and (\ref{eq:flat-rho2}) we have calculated the amount of non-gaussianity in two different situations: $(1)$ the curvaton decays into radiation and matter
with constant couplings, \ie $f_{\gamma}(N)=1$, $f_{m}(N)=1$, and $(2)$ $f_{\gamma}(N)=1$ and the decay of the curvaton field into the matter component has explicit time dependence $f_{m}(N)$.
We take $f_{m}(N)$ to be a continuous function with $f_{m}'(N)=0$ at $N=0$.

\subsection{Conserved quantities}

The authors of \cite{Gupta:2003jc} noticed that the curvature perturbation $\zeta^{(1)}_{\textrm{comp}}$, which is related to a scaled matterlike fluid component 
\begin{equation}
\rho_{\textrm{comp}}=\rho_{m}+\frac{\Gamma_{m}}{\Gamma_{\gamma}+\Gamma_{m}}\rho_{\sigma},
\end{equation}
is a conserved quantity in the three-fluid model because
\begin{equation}
Q_{\textrm{comp}}=Q_{m}+\frac{\Gamma_{m}}{\Gamma_{\gamma}+\Gamma_{m}}Q_{\sigma}=0
\end{equation}
and $\delta Q^{(1)}_{\textrm{comp}}=0$. This result can be generalised to all orders since $\delta Q^{(i)}_{\textrm{comp}}=0$
and from equation (\ref{eq:zeta-eq-2}) we can easily see that this is equal to $\zeta^{(2)}_{\textrm{comp}}{}'=0$, \ie $\zeta^{(2)}_{\textrm{comp}}$ is a conserved quantity. If we now include the possibility of time dependent interactions as in eq. (\ref{eq:interact}), $\zeta^{(1)}_{\textrm{comp}}$ is however no longer conserved because \begin{equation}
Q_{\textrm{comp}}=\frac{\Gamma_m \Gamma_{\gamma}f'_{m}(N)}{ (\Gamma_{\gamma} + \Gamma_m f_{m}(N))^2}\rho_{\sigma}\neq 0.
\end{equation}

Now in the both situations initially $\rho_{m}=0$, $\zeta^{(1)}_{\sigma,\textrm{in}}=1$ and $\zeta^{(1)}_{m,\textrm{in}}=0$ and we can see from equation (\ref{eq:zetas-flat}) that
\begin{equation}
\zeta^{(1)}_{\textrm{comp,in}}=-\frac{\delta\rho^{(1)}_{m}+\frac{\Gamma_{m}f_m(N)}{\Gamma_{\gamma}+\Gamma_{m}f_m(N)}\delta\rho^{(1)}_{\sigma}}{-3(\rho_{m}+\frac{\Gamma_{m}f_m(N)}{\Gamma_{\gamma}+\Gamma_{m}f_m(N)}\rho_{\sigma})}\Big|_{N=0} = -\frac{\frac{\Gamma_{m}f_m(N)}{\Gamma_{\gamma}+\Gamma_{m}f_m(N)}\delta\rho^{(1)}_{\sigma}}{-3\frac{\Gamma_{m}f_m(N)}{\Gamma_{\gamma}+\Gamma_{m}f_m(N)}\rho_{\sigma}}\Big|_{N=0}=-\frac{\rho_{\sigma}'}{3\rho_{\sigma}}\zeta^{(1)}_{\sigma,\textrm{in}}\Big|_{N=0}=-\frac{\rho_{\sigma}'}{3\rho_{\sigma}}\Big|_{N=0}\simeq 1,
\end{equation}
where we have used the fact that $\Gamma_{\sigma}f(N)/H\Big|_{N=0}\ll1$ in our calculations. Once the curvaton has decayed completely, $\rho_{\sigma}|_{Dec}\ll\rho_{m}|_{Dec}$ and $\zeta^{(i)}_{m}|_{Dec}\simeq\zeta^{(i)}_{\textrm{comp}}$. Now if $f_m(N)=1$, $\zeta^{(i)}_{\textrm{comp}}$ are conserved and therefore $\zeta^{(i)}_{\sigma,\textrm{in}}=\zeta^{(i)}_{m}|_{Dec}$. In terms of $r_1$ and $r_2$ this means that $r_1=1$ and $r_2=1/2$ which limits the possible values of the resulting non-gaussianity parameter (\ref{eq:fNL}) significantly.

\section{Numerical results}

\subsection{Continuous interactions, $f_{\gamma}(N)=1$, $f_{m}(N)=1$}

\begin{figure}[tbh]
\subfigure[]{\label{fig1a}\includegraphics[width=0.45\columnwidth]{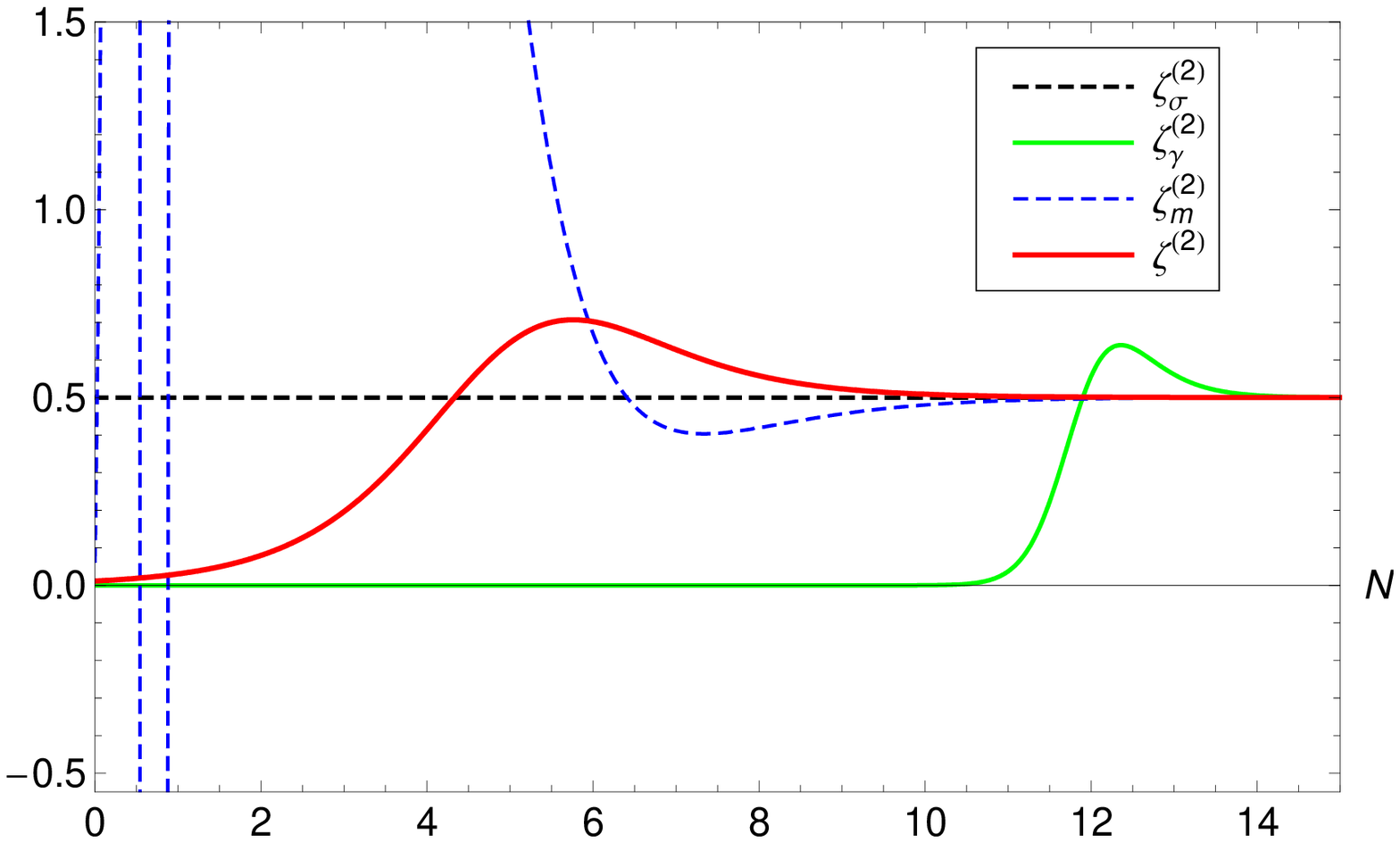}}
\quad
\subfigure[]{\label{fig1b}\includegraphics[width=0.45\columnwidth]{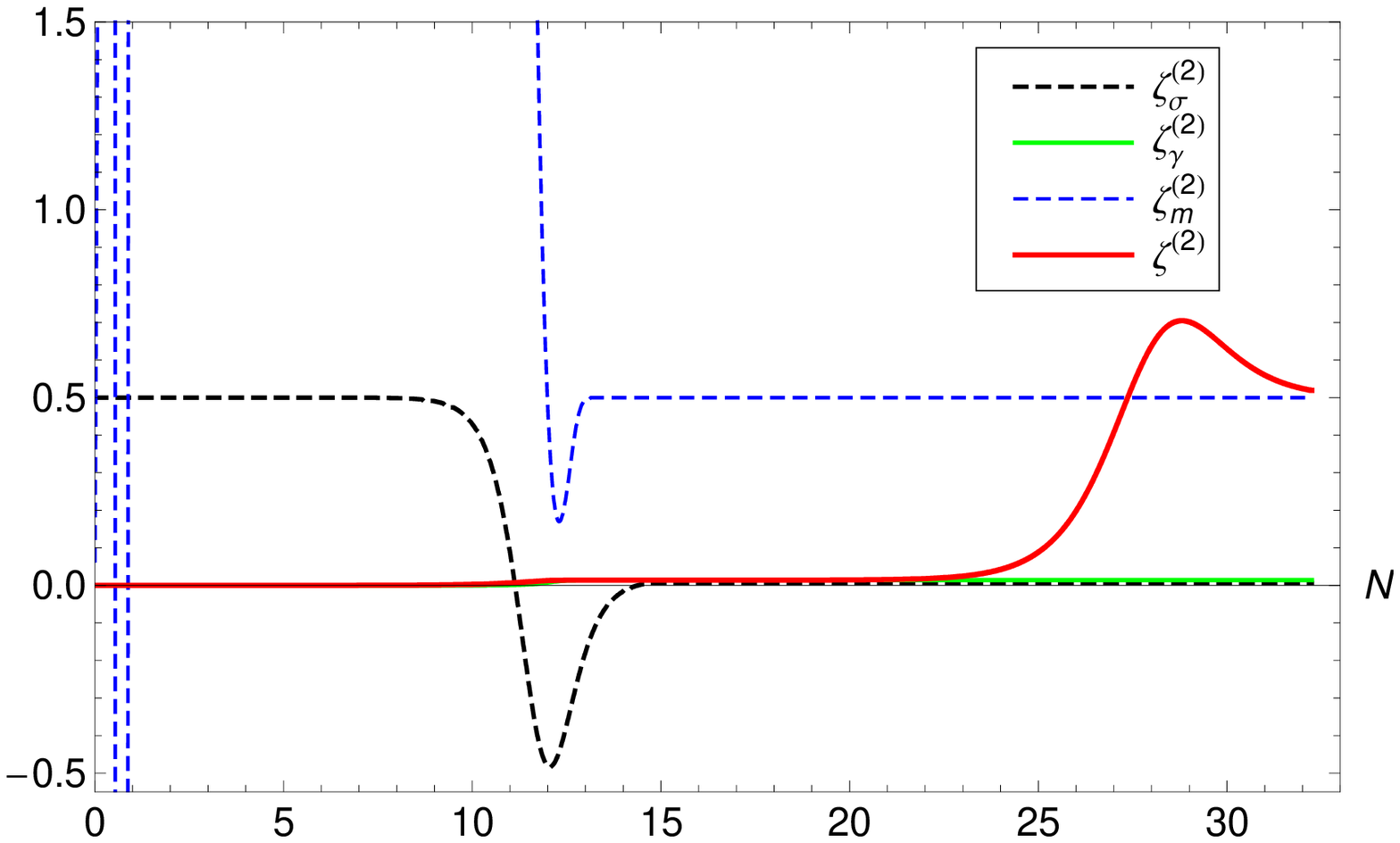}}
\caption{Evolution of the second order curvature perturbations in the curvaton-radiation-matter system with  $\Gamma_{\gamma}=10^{-10}$ and $\Gamma_{m}=10^{-15}$, when (a) $\Omega_{\sigma0}=10^{-2}$ and (b) $\Omega_{\sigma0}=10^{-7}$. Smaller initial curvaton density leads to a larger isocurvature perturbation.}
\end{figure}

The continuous case was first presented in \cite{Gupta:2003jc} whereas in \cite{Multamaki:2007hv} we studied the 
parameter space in more detail using physically motivated constraints. We also pointed out that the first order perturbation theory was not able to give limits on generated non-gaussianity. We now apply the second order theory 
results to calculate the amount of non-gaussianity. 
We have now
$\zeta^{(1)}_{m}|_{\textrm{Dec}}=\zeta^{(1)}_{\sigma,\textrm{in}}=1$ and $\zeta^{(2)}_{m}|_{\textrm{Dec}}=\zeta^{(2)}_{\sigma,\textrm{in}}=1/2$. 
The evolution of $\zeta^{(1)}_{\gamma}$ was explained in detail in \cite{Gupta:2003jc}. If the curvaton fluid begins to dominate the system before it decays almost all of the radiation originates from the curvaton and $\zeta^{(1)}_{\gamma}|_{\textrm{Dec}}\simeq\zeta^{(1)}_{\gamma,\textrm{in}}=1$. The same reasoning also applies at second order and during decoupling $\zeta^{(2)}_{\gamma}|_{\textrm{Dec}}\simeq\zeta^{(2)}_{\gamma,\textrm{in}}=1/2$ \ie in terms of $q_1$ and $q_2$, $q_1=1$ and $q_2=1/2$ and the system is adiabatic. If the curvaton decays before it begins to dominate the system, the curvature perturbations of the radiation fluid remain smaller and therefore lead to both adiabatic and isocurvature perturbations. This behaviour can be seen in the figures \ref{fig1a}-\ref{fig1b} where we have plotted the evolution of the second order perturbations $\zeta^{(2)}_{i}$ $i=\sigma,\gamma,m$. This also shows that the parameters $q_1$ and $q_2$ are not independent \ie a small absolute value of $q_1$ means that $|q_2|$ is also small whereas in the opposite case $q_2=1/2$ when $q_1=1$.


In terms of the non-gaussianity parameter the above reasoning means that eq. (\ref{eq:fNL}) can be written in the form
\begin{equation}\label{anafnl}
f_{NL}\approx\frac{25(q_2-q_1^2)-60q_1+81}{6(6-5q_1)}.
\end{equation}
From this we can see that an adiabatic system, \ie $q_1=1, q_2=1/2$, gives $f_{NL}=17/12\approx 1.42$. The maximum value $f_{NL}=2.73$ corresponds to values $q_1=0.438$, $q_2 = 0.5$ \ie the system is non-adiabatic.

For different interaction strengths the results are similar and in good agreement with the analytical
approximation: the maximum values of non-gaussianity are close to the expected value and $f_{NL}$ decreases as the system becomes more adiabatic. In a previous paper we studied the same system in detail \cite{Multamaki:2007hv} and found that the system becomes adiabatic as the interaction strengths decrease, hence weaker interaction strengths lead
to less non-gaussianity.

\subsection{Time dependent interactions, $f_{\gamma}(N)=1$, $f_{m}(N)=\theta(N-N_*)$}


The time dependent scenario was presented in detail in \cite{Multamaki:2007hv}, where we found only small differences compared to the continous case.  Our choice for the interaction function $f(N)$ is $f(N)=(\tanh((N-N_*)/\tau)+1)/2$ where $\tau=10^-5$. Major alteration comes from the change of $H_{0}\rightarrow H_{*}<H_{0}$, which shifts different regions upward compared to the time independent scenario. 
We find a similar behaviour here.
In terms of the curvature perturbation $\zeta_{\textrm{comp}}$ it is no longer conserved because $Q_{\textrm{comp}} \neq 0$. However since the function $f(N)$ reaches value $1$ very quickly, $f'(N)=0$ is true almost everywhere and $\zeta_{\textrm{comp}}$ is almost conserved. 
The value of the non-gaussianity produced in this scenario is thus very similar to the constant interaction case.

\section{Discussion and conclusions}

In this paper we have studied the generation of non-gaussianity in the three fluid model of curvaton decay by means of second order perturbation theory. In the first part of this paper we introduced general formulae of perturbation theory, different gauge conditions and concluded it with a derivation of a general formula for the Sachs-Wolfe effect on large scales.
It includes the possibility of a non-adiabatic final state and simplifies to the adiabatic formulas presented previously in \cite{Bartolo:2004if}. In the second part of this paper we applied this formula to the three fluid model of a curvaton decay and studied the generation of non-gaussianity in the temperature anisotropy of the CMB.
We find that in general the amount of non-gaussianity produced in this scenario is small, $f_{NL}\sim \mathcal{O}(1)$
both for constant and dynamical interaction strengths between the fluids.

Our results indicate that the three fluid model leads to less non-gaussianity than the two fluid one. This is especially true if the curvaton field fails to dominate the system which leads to large non-gaussianity in the two fluid model because
then $f_{NL}\simeq-5/(4r)$ where $r$ is small \cite{Bartolo:2004if, Komatsu:2003fd}. 
If the detected non-gaussianity is small one cannot, however, conclude that the three fluid model is responsible for it
since such an observation is in agreement with the standard inflationary scenario $f_{NL}=-1/2$ \cite{Bartolo:2004if,Bartolo:2003gh}.
One way to break this degeneracy is to compare the generation of gravity waves in the curvaton scenario \cite{Bartolo:2007vp} with the standard inflation which usually produces much higher level of gravity waves.
Another approach is to use the isocurvature: A non-adiabatic universe with a small amount of non-gaussianity would indicate the three fluid curvaton
model as a natural candidate.

\subsection*{Acknowledgments}
This project has been partly funded by the Academy of Finland project no. 8111953.
TM and JS are supported by the Academy of Finland.

\newpage


\end{document}